\documentclass[
  twocolumn,  
  amsmath, amssymb,
  superscriptaddress,floatfix
]{revtex4-2}

\usepackage{float}
\makeatletter
\let\newfloat\newfloat@ltx
\makeatother
\usepackage{algorithm}
\usepackage{algpseudocode}
\usepackage[utf8]{inputenc}
\usepackage[T1]{fontenc}
\usepackage{graphicx}
\usepackage{dcolumn}
\usepackage{bm}
\usepackage{mathrsfs}
\usepackage{siunitx}
\usepackage{xcolor}
\usepackage{hyperref}
\usepackage{microtype}
\usepackage{physics}
\usepackage{braket}
\usepackage{dblfloatfix}  
\usepackage{microtype}

\begin{document}
\sloppy 

\title{\texorpdfstring{Hardware Co-Designed Optimal Control for Programmable Atomic \\ Quantum Processors via Reinforcement Learning}{Hardware Co-Designed Quantum Optimal Control for Programmable Neutral Atom Processors via Reinforcement Learning}}

\author{Qian Ding}
\email{dingq30@mit.edu}
\affiliation{Research Laboratory of Electronics, Massachusetts Institute of Technology, Cambridge, MA 02139}

\author{Dirk Englund}
\affiliation{Research Laboratory of Electronics, Massachusetts Institute of Technology, Cambridge, MA 02139}
\affiliation{Department of Electrical Engineering and Computer Science, Massachusetts Institute of Technology, Cambridge, MA 02139}

\begin{abstract}
\vspace{0.2cm}
\centering
\parbox{0.9\textwidth}{
\setlength{\parindent}{2em}
\noindent Developing scalable, fault-tolerant atomic quantum processors requires precise control over large arrays of optical beams. This remains a major challenge due to inherent imperfections in classical control hardware, such as inter-channel crosstalk and beam leakage. In this work, we introduce a hardware co-designed intelligent quantum control framework to address these limitations. We construct a mathematical model of the photonic control hardware, integrate it into the quantum optimal control (QOC) framework, and apply reinforcement learning (RL) techniques to discover optimal control strategies.
We demonstrate that the proposed framework enables robust, high-fidelity parallel single-qubit gate operations under realistic control conditions, where each atom is individually addressed by an optical beam. Specifically, we implement and benchmark three optimization strategies: a classical hybrid Self-Adaptive Differential Evolution–Adam (SADE-Adam) optimizer, a conventional RL approach based on Proximal Policy Optimization (PPO), and a novel end-to-end differentiable RL method. Using SADE-Adam as a baseline, we find that while PPO performance degrades as system complexity increases, the end-to-end differentiable RL consistently achieves gate fidelities above 99.9$\%$, exhibits faster convergence, and maintains robustness under varied channel crosstalk strength and randomized dynamic control imperfections.
}
\end{abstract}

\maketitle
\section{Introduction}
\noindent Quantum computing has the potential to revolutionize fields such as cryptography, optimization, chemistry, and materials science by efficiently solving problems that are intractable for classical computers~\cite{Preskill2018QuantumComputing, Montanaro2016QuantumAlgorithms, Arute2019QuantumSupremacy, Bharti2022Noisy}. However, building scalable and fault-tolerant quantum processors remains extremely challenging due to the stringent requirement for high-fidelity gate operations. These requirements, in turn, demand precise and robust control of quantum systems~\cite{Gambetta2017Building, google2023, Atom2016Programmable}.
Neutral atom quantum processors are promising candidates for scalable quantum computing given their excellent coherence properties, inherent scalability, and flexible reconfigurability through optical control~\cite{Henriet2020QuantumComputing, Morgado2021Quantum, DB2022}. Recent experimental advances have demonstrated large-scale atom arrays capable of high-fidelity logical operations and efficient entanglement generation~\cite{Ebadi2021QuantumSimulation, DB2024}. However, as these systems scale, the challenge of generating and delivering precise control pulses to individual atoms becomes more pronounced, especially in the presence of hardware imperfections that cause crosstalk between control channels.
Quantum optimal control (QOC) provides a powerful theoretical framework for designing control pulses that drive high-fidelity quantum operations~\cite{Khaneja2001, Krotov2012, Caneva2011CRAB}. More recently, reinforcement learning (RL) has emerged as a promising approach in QOC, enabling adaptive control strategies that do not require full knowledge of system dynamics~\cite{Bukov2018Reinforcement, Yang2020RLforQOC, Porotti2022DeepRL, VVS2022}. Yet, the integration of QOC theory with realistic quantum control hardware remains largely unexplored~\cite{Koch2022QOCReview}, particularly the potential of RL-based QOC methods to mitigate control challenges induced by non-ideal hardware systems.

\medskip

\noindent To bridge this gap, we propose a hardware co-designed QOC framework driven by advanced classical and RL-based optimization algorithms. This framework supports direct integration of detailed mathematical models of the control hardware, including photonic integrated circuits (PICs) and spatial light modulators (SLMs), into the QOC formalism. By explicitly modeling physical imperfections like inter-channel crosstalk and beam leakage, the RL agent learns to program the hardware to implement robust, high-fidelity quantum gate operations. However, embedding hardware models into the QOC problem significantly increases its complexity, making the optimization landscape more non-convex and high-dimensional than in idealized pulse design. This motivates the adoption of more powerful optimization strategies beyond commonly used techniques like GRAPE or Krotov.
To this end, we implement and evaluate three distinct optimization methods: (1) a hybrid classical approach combining Self-Adaptive Differential Evolution and Adam (SADE-Adam), (2) a conventional RL method based on Proximal Policy Optimization (PPO), and (3) a novel end-to-end differentiable RL-based optimizer.
As a proof of concept, we apply this framework to implement parallel single-qubit gates on a neutral atom processor, controlled by realistic hardware that includes a programmable PIC and SLM. Ideally, each atom is addressed by an independent optical beam to realize the target gate; however, in practice, non-idealities such as waveguide-induced crosstalk and beam leakage from imperfect steering lead to errors in gate execution. We benchmark the performance of the three optimization strategies using the classical SADE-Adam optimizer as a baseline and compare it with the PPO-based and end-to-end differentiable RL approaches.
Our results show that the end-to-end differentiable RL method begins to outperform the others as the number of
\newpage
\onecolumngrid
\begin{figure*}[htbp]
    \includegraphics[width=1.0\textwidth]{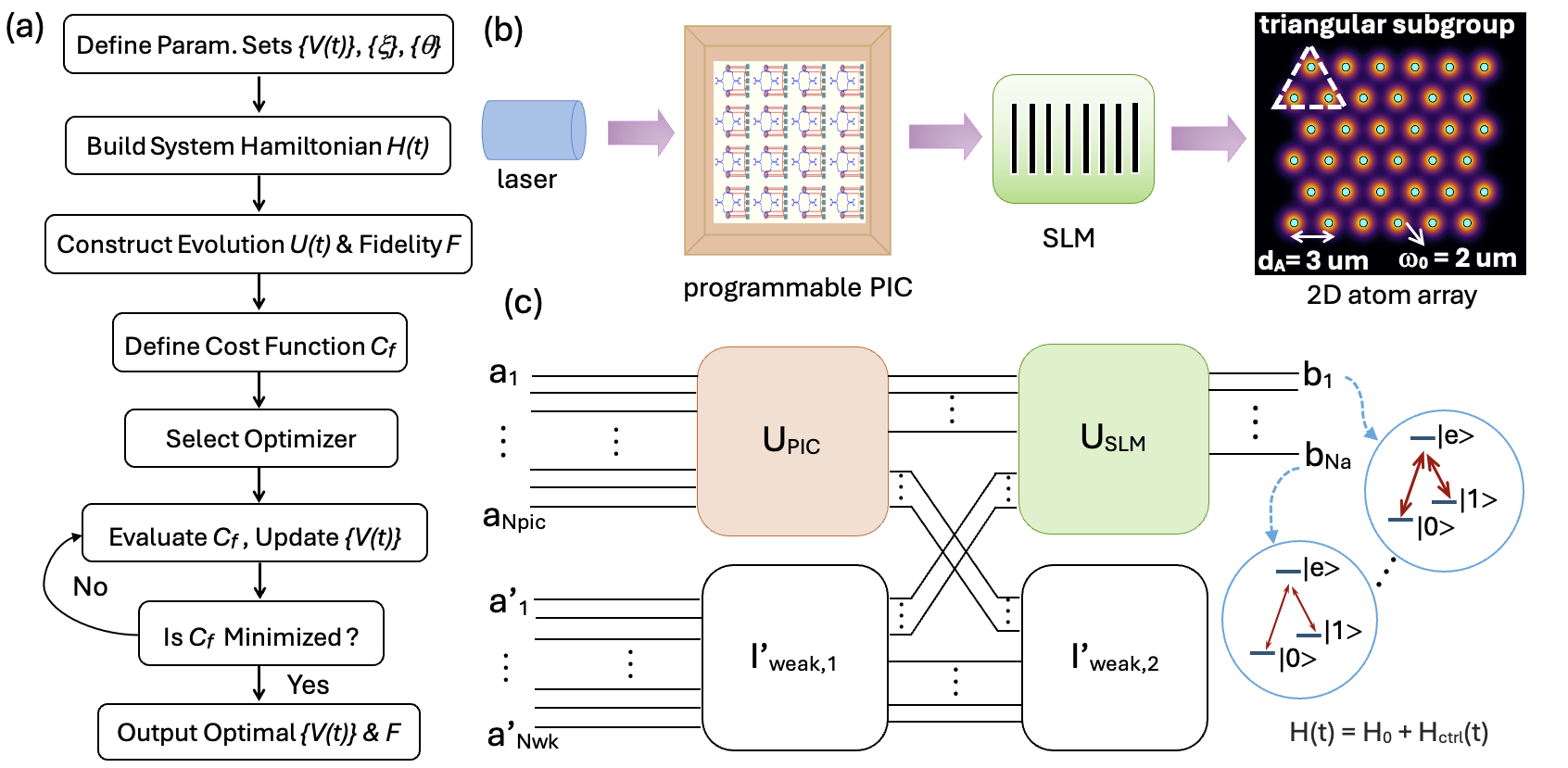}
    \caption{(a) Workflow of the implemented hardware co-designed QOC framework. The process starts with defining system parameters and initial control pulses, then constructing mathematical model of control hardware and further system Hamiltonian, further simulating quantum evolution and computing cost function, and iteratively optimizing control pulses until the cost function is minimized under given constraints.
    (b) A conceptual sketch of the control system for neutral atom quantum processors: laser beams first get modulated via tunable units in programmable PICs and then dynamically steered by a SLM onto a neutral atom array (assuming a triangular lattice here) to implement target gate operations. 
    (c) Mathematical model of the control hardware system using unitary transformation representation. The input modes \( \{a_1, \dots, a_{N_{\mathrm{pic}}}\} \) are modes coupled from free space into the PIC, which is modeled as a unitary matrix \( U_{\mathrm{PIC}} \), incorporating transformations induced by both programmed modulations and unintentional inter-channel crosstalk. Weak scattering effects are modeled by a slightly perturbed identity matrix \( I'_{\mathrm{weak,1}} \). The output modes from the PIC further get transformed by a SLM, represented by another unitary matrix \( U_{\mathrm{SLM}} \). Weak scattering effects in this stage are captured by \( I'_{\mathrm{weak,2}} \). The output modes \( \{b_1, \dots, b_{N_a}\} \) are steered to the target atoms to implement a desired gate operation by programming the control Hamiltonian. 
}\label{fig:physical_model}
\end{figure*}
\twocolumngrid

\noindent simultaneously applied single-qubit gates increases. It achieves faster convergence, consistently maintains gate fidelities above 99.9\%, and remains robust even under randomized dynamic control imperfections.
The remainder of this paper is organized as follows: Sec.~\ref{sec:math_formulation} presents the mathematical formulation of the hardware co-designed QOC problem and describes the physical models of the control hardware and atomic qubit system. Sec.~\ref{sec:methods} details the implementation of the three optimization strategies. Sec.~\ref{sec:results} reports numerical results and benchmarking comparisons. Concluding remarks and future directions are discussed in Sec.~\ref{sec:conclusion}.

\section{Mathematical Models}
\label{sec:math_formulation}
\noindent In this section, we first describe how we mathematically formulate the hardware integrated quantum optimal control problem, then we present detailed models to describe the physics of the programmable photonic control hardware considering practical inter-channel crosstalk and the neutral atom array quantum system.

\subsection{Formulating Co-designed Optimization Problem}
\noindent We formalize the hardware co-designed QOC problem as a constrained optimization task, which aims to find the optimal control pulses applied to programmable units in the hardware to implement high-fidelity target gate operations. The entire workflow the co-designed QOC optimizer is illustrated in Fig.~\ref{fig:physical_model} (a), with details  explained in the following. 
A mathematical model $\mathscr{M}(\{\theta\}, \{V(t)\})$ of the control hardware describes how the input laser fields \(\{a_{in}\}\) are mapped to a set of dynamically modulated output fields $\{b_{out}(t)\}$ that address the involved atom qubits for gate implementation.
Here, $\{\theta\}$ represent a set of design parameters characterizing the control hardware \(\{V(t)\}\) are the electrical signals applied to the programmable units in PIC for field modulation.
Incorporating realistic hardware imperfections such as inter-channel crosstalk and beam leakage can be directly captured in this mathematical map.
The output fields $\{b_{out}(t)\}$ further enter the control Hamiltonian $H_{control}(t)$ that determines the system evolution and final gate fidelity. The total system Hamiltonian is given as:
\begin{equation}
    H(t) = H_0(\{\xi\}) + H_{\text{control}}(
    \{b_{out}(t)\}), 
\end{equation}
where \(H_0(\{\xi\})\) represents the static drift term that models the atom qubit system with a set of parameter \(\{\xi\}\) describing the properties of atom qubits. \(H_{\text{control}}(\{b_{out}(t)\})\) denotes the control Hamiltonian model the interaction between atoms and a set of modulated control pulses \(\{b_{out}(t)\}\).
The goal is to optimize control signals \(\{V(t)\}\) to maximize the fidelity of a target quantum gate with a given gate time considering the practical hardware. We formulate the co-designed QOC problem as follows:
\begin{align}
\text{Minimize:}\quad & \mathcal{C}_f(\{V(t)\}) = 1 - F(U(T_g)), \label{eq:optimization_problem}\\ \nonumber \\
\text{subject to:}\quad & \frac{dU(t)}{dt} = -i\,H(t)\,U(t),\quad U(0) = I, \\ \nonumber \\
& V_{\min} \leq V(t) \leq V_{\max},\quad \forall t \in [0, T_g], \\ \nonumber \\
& \text{static parameter sets}\quad \{\theta, \xi\}.
\end{align}
\(F(U(T_g))\) quantifies the gate fidelity for a target operation \(U_{\text{target}}\) with a given gate time \(T_g\) as:
\begin{equation}
    F(U(T_g)) = \frac{1}{d^2}\, \left|\text{Tr}(U^\dagger_{\text{target}} U(T_g)) \right|^2,
    \label{gateF}
\end{equation}
with \(d\) denoting the Hilbert-space dimension of the quantum system. 
The actual implemented system evolution \(U(T_g)\) for the given gate time duration \(T_g\) is given by:
\begin{equation}
    U(T_g) = \mathcal{T}\exp\left(-i\int_0^{T_g}H(t)dt\right),
\end{equation}
where \(\mathcal{T}\) is the time-ordering operator.
Additionally, the control signals \(\{V(t)\}\) are subject to practical hardware constraints, captured by:
\begin{equation}
    V_{\min} \leq V(t) \leq V_{\max},
\end{equation}
reflecting realistic power limitations inherent to photonic control hardware.
In our formulation, the quantum system parameters \(\{\xi\}\), such as atomic transition frequencies and dipole moments, and photonic hardware parameters \(\{\theta\}\), including waveguide dimensions, material properties, and optical components, are assumed to remain fixed. These parameters define how the control signals affect quantum evolution, thereby shaping the optimization landscape. 
To effectively solve this optimization problem, we investigate three different approaches, with detailed implementation described in Sec.~\ref{sec:methods}.
Next, we explain how to construct the mathematical model of photonic control hardware \(\mathscr{M}\), including inter-channel crosstalk induced by waveguide coupling in practical PICs, and the quantum model for the atom qubit system using Jaynes-Cummings Hamiltonian.
\subsection{Physical System Model}
\label{sec:phys_model}
\noindent An overview of the full physical control architecture is shown in Fig.~\ref{fig:physical_model}(b). The system consists of a programmable photonic integrated circuit (PIC) that modulates the input laser fields, followed by a spatial light modulator (SLM) that dynamically steers the modulated fields to target atom qubits. These shaped optical fields then interact with the atoms to realize the desired quantum gate operations. This photonic control scheme has been experimentally demonstrated in recent works for both neutral atom and color center platforms~\cite{am2023, ic2025}.
In what follows, we present a mathematical model of the full control system, capturing the two-stage optical field manipulation pipeline: first by the programmable PIC, and then by the SLM. The PIC model incorporates both intentional modulation via on-chip components and unintentional inter-channel crosstalk due to waveguide coupling. The SLM is modeled as a dynamic beam-steering unit with programmable phase control. We then describe the atom qubit system, including its time-dependent Hamiltonian and the gate fidelity metric used to quantify control performance.\\

\begin{figure*}[htbp]
    \includegraphics[width=0.99\textwidth]{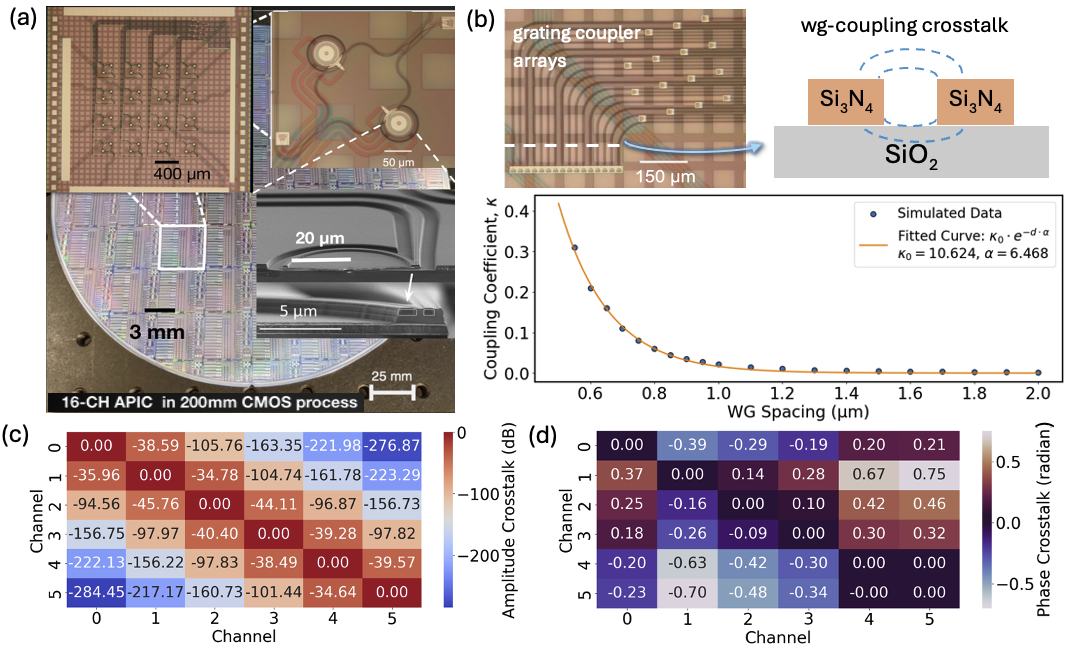}
    \caption{
    (a) A 16-channel programmable atomic PIC control hardware fabricated using a 200 mm CMOS process~\cite{Dong2022}. Insets show SEM images of the piezo-actuated Si$_3$N$_4$ dual-ring Mach-Zehnder modulator (DRMZM).
    (b) Simulated inter-channel crosstalk arising from evanescent waveguide coupling in the PIC. Dotted curves represent data from 2D FDTD simulations using \textit{FlexCompute Tidy3D}, assuming refractive indices of 2.0255 for Si$_3$N$_4$ and 1.45 for SiO$_2$ at a wavelength of 780 nm. Coupling coefficient \(\kappa_0 = 10.145\) and decay factor \(\alpha = 6.934\) are extracted from the fitted solid curve.
    (c, d) Heatmaps showing the amplitude (c) and phase (d) crosstalk matrices induced by waveguide coupling in a six-channel PIC. The matrices are generated using the fitted model from (b), assuming a coupling length of 600~\(\mu\)m and a channel pitch of 1.0~\(\mu\)m.}
    \label{fig:ct_model}
\end{figure*}

\paragraph {Control Hardware Model}
\label{subsec:hardware}
\noindent The control system is mathematically modeled using unitary matrix representations, as illustrated in Fig.~\ref{fig:physical_model} (c). The unitary matrices \(U_{\mathrm{PIC}}\) and \(U_{\mathrm{SLM}}\) describe the transformation of field modes applied via the PIC and SLM, and the slightly perturbed identity matrix \(I'_{weak,1(2)}\) represent the weak scattering effects. The input optical fields, denoted as \( \{a_{\mathrm{in}} \}\) in the following, consists of primary optical modes coupled into the PICs, noted as modes \( \{a_1, \dots, a_{N_{\mathrm{pic}}}\} \), and weakly scattered modes \( \{a'_1, \dots, a'_{N_{\mathrm{wk}}}\} \).
The output fields reaching target atom sites, noted as \(\{b_{out}(t)\}\) in the following, are given by modes \( \{b_1, \dots, b_{N_{\mathrm{a}}}\} \), with \(N_a\) being the number of target atoms.
The mapping from input optical field modes \( \{a_{\mathrm{in}} \}\) to output field modes \(\{b_{out}(t)\}\) reaching target atom sites contains two sequential stages. In the first stage, the input modes are modulated by the programmable PIC, which also introduces inter-channel crosstalk. The intermediate output field modes \(\{b_{\mathrm{out}}^{\mathrm{PIC}}(t)\}\) from this stage serve as inputs to the second stage transformation applied by the SLM. 
Mathematically, the transformation at the first stage is described as:
\begin{equation}
\{b_{\mathrm{out}}^{\mathrm{PIC}}(t)\} = \Bigl[ U_{\mathrm{PIC}}(\{V(t)\}) \otimes I'_{\mathrm{weak,1}} \Bigr] \{a_{\mathrm{in}}\},
\end{equation}
where \( U_{\mathrm{PIC}}(\{V(t)\}) \) is a tridiagonal matrix capturing the effects of programmable field modulation (on-diagonal elements) and inter-channel crosstalk (off-diagonal elements) in the PIC. 
In each channel, the programmed field modulation is represented by a transfer function \(T\{V(t)\}\) derived based on the physics model of the control unit, which is a dual-ring Mach-Zehnder modulator (DRMZM) made of \(Si_3N_4\), with its SEM image given on the upper right in Fig.~\ref{fig:ct_model} (a).   
The field modulations via programmed control pulses \(\{V(t)\}\) applied to DRMZMs rely on tunable \\
\newpage
\onecolumngrid
\twocolumngrid

\noindent piezo-opto-mechanical actuation ~\cite{Eichenfield2019}. Details of device structure and transfer function model of the DRMZM can be found in Ref.~\cite{am2023}. 
The inter-channel crosstalk effect induced by evanescent coupling between waveguides in channels \(m\) and \(n\) is captured using a coupling coefficient \(C_{m,n} = A_{m,n} e^{i \theta_{m,n}}\). According to coupled mode theory ~\cite{CMT} under assumption of a symmetric directional coupler, we set amplitude \(A_{m,n}\) as \(\left|\sin\left(\kappa_{\text{eff},m,n}\, L_{m,n}\right)\right|\) and phase \(\theta_{m,n}\) as \(-\pi/2\). The effective coupling factor \(\kappa_{\text{eff},m,n}\) is calculated as \( \sqrt{\kappa_{m,n}^2 + \Delta\beta^2}\), with the nominal coupling coefficient \(\kappa_{m,n} = \kappa_0 \exp\left(-\alpha\, d_{m,n}\right)\) and phase matching \(\Delta\beta = \frac{\beta_m - \beta_n}{2}\). Here, the propagation constant \(\beta_i\) for channel \(i\) is given by \(\frac{2\pi}{\lambda_0}\, n_{\text{eff},i}\) with \(n_{eff,i}\) being the effective refractive index of the waveguide in channel \(i\) and \(\lambda_0\) being the wavelength of the propagating field , which is \(780\,nm\) in our case. For \(\kappa_{m,n}\), the prefactor \(\kappa_0\) is the base coupling coefficient for nearest-neighboring waveguides, \(\alpha\) is the decay factor that governs how quickly the coupling strength falls off with distance \(d_{m,n}\) between channels \(m\) and \(n\), and \(L_{m,n}\) is the coupling length between the involved waveguides. 
These two parameters, \(\kappa_0\) and \(\alpha
\), can be obtained by fitting to 2D finite-difference time-domain (FDTD) simulation results obtained for neighboring \(Si_3N_4\) waveguides, as seen in Fig.~\ref{fig:ct_model} (b).
To account for practical fabrication variations of PICs in the foundry, the values of \(d_{m,n}\) and \(L_{m,n}\) are determined by combining a deterministic component with random variations. Specifically, the distance between channels is calculated as \(d_{m,n} = |m - n| \times d_0 + \delta_d\), where \(d_0\) is the nominal channel pitch and \(\delta_d\) is a small random variation uniformly sampled from a specified range. Similarly, the coupling length between channels is computed as \(L_{m,n} = L_0 \times s^{|m - n|} + \delta_L\), where \(L_0\) is the nominal coupling length for waveguides in nearest-neighbor channels, \(s\) is a scaling factor adjusting the nominal coupling length with increasing channel separation, and \(\delta_L\) is a random perturbation uniformly sampled from a given range. As an example, Fig.~\ref{fig:ct_model}(c) and (d) show the calculated amplitude and phase crosstalk matrices for a 6-channel photonic integrated circuit (PIC), where the nominal coupling length is \(600\,\mu\text{m}\) (with \(s\) set to 1.1 and \(\delta_L\) uniformly sampled from \([-60\,\mu\text{m},\,60\,\mu\text{m}]\)) and the channel spacing is \(1.0\,\mu\text{m}\) (with \(\delta_d\) uniformly sampled from \([-0.1\,\mu\text{m},\,0.1\,\mu\text{m}]\)).
Given the on-diagonal evolution represented by \(T\{V(t)\}\), which corresponds to the model of channel-wise modulator unit, and the off-diagonal crosstalk coefficients \(C_{m,n}\), the final unitary matrix modeling the PIC is constructed as:
\[
\bigl[ U_{\mathrm{PIC}}(t) \bigr]_{m,n} = \begin{cases}
T\{V(t)\}, & \text{if } m = n, \\
C_{m,n}, & \text{if } m \neq n.
\end{cases}
\]
Next, we construct the transformation matrix of the second stage involving the SLM, which is given by:
\begin{equation}
b_{\mathrm{out}}(t) = \Bigl[ U_{\mathrm{SLM}}(t) \otimes I'_{\mathrm{weak,2}}(t) \Bigr] b_{\mathrm{out}}^{\mathrm{PIC}}(t).
\end{equation}
Here, \( U_{\mathrm{SLM}}(t) \) is a diagonal unitary matrix encoding pixel-wise modulations applied by the SLM.
In this work, we consider the manipulation of SLM as static and it has a fixed amplitude and phase modulation in each channel.
Among the multiple output modes in \( b_{\mathrm{out}}(t) \), a subset \( \{b_1, \dots, b_{N_a}\} \) corresponds to the optical fields that reach and interact with the target atoms. These field modes form structured optical wavefronts as they projected onto the plane of a given 2D atom array. Each optical field is assumed to have a Laguerre-Gaussian (LG$_{00}$) mode profile with a beam waist of \(2\,\mu m\), when constructing the intensity and phase distribution of the field on the 2D plane. On this plane, the atoms are arranged in equilateral triangular sublattices with an inter-atomic spacing of \(3 \,\mu m \), as illustrated in Fig.~\ref{fig:physical_model} (b). 
The chosen configuration of atom array and beam waist here allows us naturally include the beam leakage effect.
The complex-valued field extracted at each target atom site drives the atom-field interaction, generating an effective Hamiltonian that governs the quantum evolution and further determines gate fidelity, as described below.\\

\paragraph{Qubit System Model}
\label{subsec:qubitmodel}
The system Hamiltonian consists of both drift and control components:
\begin{equation}
{H}(t) = {H}_0 + {H}_\mathrm{control}(t).
\end{equation}
The drift Hamiltonian describes the free evolution of the qubits and the quantized optical field:
\begin{equation}
{H}_0 = \frac{1}{2} \omega_0 \sum_{j=1}^{N_a} \hat{\sigma}_z^{(j)} + \omega_r \hat{b}^\dagger \hat{b},
\end{equation}
where  \( \omega_{0(r)}\) is the frequency of atomic transition (the quantized field mode) and \( N_a \) is the number of atom qubits.
The control Hamiltonian describes the light-matter interaction using Jaynes Cummings model \cite{JC1963}:
\begin{equation}
{H}_\mathrm{control}(t) = \sum_{j=1}^{N_a} \left[ g_j(t) \hat{\sigma}_+^{(j)} \hat{b} + g_j^*(t) \hat{\sigma}_-^{(j)} \hat{b}^\dagger \right].
\end{equation}
The interaction between the optical fields and qubits is mediated by a Raman process \cite{raman2022}.
The effective interaction strength \( g_j(t) \) is given by
\(\frac{\Omega_1 \Omega_2^*}{2\Delta}\),
where the Rabi frequencies of the two-photon process are
$\Omega_{1,2} = {\mu_{1,2e} * E_j(t)}/{\hbar}$, with
\( \mu_{1,2e} \) being the dipole moments associated with the atomic transitions of typical values \(2.54 \times 10^{-29} \,\mathrm{C \cdot m} \). The detuning \(\Delta\) is set to \(1 \mathrm{GHz} \). The local electric field \( E_j(t) \) is the complex-valued field extracted at $j$-th target atom site.
Given unitary time evolution of the quantum system under the total Hamiltonian, the propagator can be computed numerically using a discrete time grid as:
\begin{equation}
U(T_g) \approx \prod_{t=1}^{t_{\mathrm{steps}}} \exp \bigl[ -i H(t) \Delta t \bigr].
\end{equation}
Here, \( t_{\mathrm{steps}} \) are the total number of discrete time steps, \( \Delta t = T_g / t_{\mathrm{steps}} \) is the discrete time step size, and \(T_g\) is a given gate operation time.
The gate fidelity is then calculated as:
\begin{equation}
F = \frac{1}{(2^{N_a})^2} \left| \mathrm{Tr} \Bigl[ U_{\mathrm{target}}^\dagger U_{\mathrm{q}}(T_g) \Bigr] \right|^2,
\label{gateF}
\end{equation}
where \( U_{\mathrm{q}}(T_g) \) is the propagator describing the evolution of qubit system obtained by tracing out the field modes in \(U(T_g)\).
For the target gate, we consider parallel single-qubit gates implemented simultaneously on \(N_a\) atoms. The target gate on each atom, denoted by \(R_j\), is randomly selected from the extended Clifford group with T-gate. Consequently, the overall target unitary is given by:
\begin{equation}
U_{\mathrm{target}} = R_1 \otimes R_2 \otimes \cdots \otimes R_{N_a}.
\end{equation}
In this work, we consider three atoms, and the construction of \(U_{\mathrm{target}}\) depends on the specific program instruction. For instance, if the instruction is to implement a randomly selected single-qubit gate on atom 1 only, then
\(
U_{\mathrm{target}} = R_1 \otimes I_2 \otimes I_3,
\)
where \(I_2\) and \(I_3\) denote the identity gates on atoms 2 and 3, respectively. If the instruction is to implement randomly selected single-qubit gates on atoms 1 and 2, then
\(
U_{\mathrm{target}} = R_1 \otimes R_2 \otimes I_3.
\)
Finally, if all three atoms receive randomly selected single-qubit gates, then
\(
U_{\mathrm{target}} = R_1 \otimes R_2 \otimes R_3.
\)
We label the three cases as the easy, intermediate, and hard tasks for our co-designed QOC problem. These  designations will be used when discussing numerical results in Sec.~\ref{sec:results}.

\section{Optimization Methods}
\label{sec:methods}
\noindent To solve the co-designed QOC optimization problem formulated in Section~\ref{sec:math_formulation}, we implement three optimizers that can automatically design control pulses to implement high-fidelity, robust parallel single-qubit gates in presence of hardware control imperfections like inter-channel crosstalk and beam leakage. Specifically, these optimizers include (i) a classical hybrid SADE-Adam optimizer, (ii) a conventional RL approach based on PPO~\cite{Schulman2017ProximalPO}, and (iii) an end-to-end differentiable RL optimizer. 
In both RL methods, the policy network is used to generate a candidate control pulse set \(\{V(t)\}\). 
All three optimization algorithms are implemented using JAX~\cite{jax2018}, allowing for automatic differentiation and accelerated computations. Implementation details of each optimizer are given in the following subsections.

\subsection{Hybrid SADE-Adam Approach}
\label{sec:sade_adam}
\noindent SADE is an evolutionary optimization algorithm that dynamically adapts its mutation and crossover rates to balance exploration and exploitation \cite{Qin2009Differential, Jiao2020Advances}, making it well-suited for global searches in high-dimensional spaces. Adam is a gradient-based optimizer that combines first- and second-moment estimates for adaptive learning rate updates \cite{Kingma2014Adam, Reddi2019Convergence}, ensuring stable and efficient fine-tuning of control parameters.
We implement a hybrid approach combining both methods to ensure fast convergence, leveraging SADE for coarse global exploration and Adam for fine-grained optimization. Specifically, we first run the SADE algorithm to find a candidate control signal set \(\{V(t)\}\) that achieves a relatively high fidelity (e.g. above 0.95). Then Adam optimizer takes this sub-optimal solution as its initial guess and search for a solution to achieve a higher fidelity above a desired threshold (e.g. 0.999). The overall procedure is summarized in Algorithm~\ref{alg:sade_adam}. Detailed implementations are given in Appendix~\ref{appendix:sade_adam}.

\begin{algorithm}
\caption{Hybrid SADE-Adam Approach}
\label{alg:sade_adam}
\begin{algorithmic}[1]
\State \textbf{Input:} Initial solution \(x_0\), population size \(P\), fidelity threshold \(F_{\text{thresh}}\), max SADE generations \(G\), Adam parameters (initial LR \(\alpha_0\), max steps \(T\))
\State \textbf{Initialize:} Population \(\{x_i\}_{i=1}^{P}\) around \(x_0\)
\For{generation \(g=1\) to \(G\)}
    \For{each candidate \(x_i\)}
        \State \textbf{Mutation:} \(m = x_a + \mu\,(x_b - x_c)\) \Comment{\(\mu \in [0.1,0.9]\)}
        \State \textbf{Crossover:} For each component, 
        \[
          u_{i,j} =
          \begin{cases}
          m_j, & \text{if } r_j < CR,\\[1mm]
          x_{i,j}, & \text{otherwise.}
          \end{cases}
        \]
        \State \textbf{Selection:} If \(f(u_i) < f(x_i)\), then set \(x_i \gets u_i\)
    \EndFor
    \State Update best candidate: \(x^* \gets \arg\min_{x_i} f(x_i)\)
    \If{Fidelity(\(x^*\)) \(\ge F_{\text{thresh}}\)}
         \State \textbf{break} from SADE loop
    \EndIf
\EndFor
\State \textbf{Switch to Adam:} Initialize \(x \gets x^*\)
\For{step \(t=1\) to \(T\)}
    \State Compute gradient \(g=\nabla f(x)\)
    \State Update: \(x \gets x - \alpha_t\,g\) \Comment{via Adam with clipping}
    \State Decay \(\alpha_t\) if no improvement
    \If{Fidelity(\(x\)) converges}
         \State \textbf{break}
    \EndIf
\EndFor
\State \textbf{Output:} Optimized solution \(x\)
\end{algorithmic}
\end{algorithm}

\subsection{PPO-Based RL Approach}
\noindent To formulate our hardware co-designed QOC problem as a RL problem, we cast the task of steering a quantum system toward a desired gate operation into the framework of PPO, which is a widely used policy-gradient algorithm designed to improve training stability while maintaining sample efficiency. Specifically, we design a custom Gymnasium environment, \textit{QOCEnv}, where the state is defined by the historical control voltages and current fidelity metrics, and the action space comprises continuous voltage adjustments that are later scaled to the physical range of \([-15\,\mathrm{V}, 15\,\mathrm{V}]\). The reward function is carefully tailored to incentivize improvements in gate fidelity via adaptive scaling:
\begin{equation}
r_t = a\, F_t^{p} + b\, \Delta F_t,
\end{equation}
where \(F_t\) denotes the fidelity at time \(t\), \(\Delta F_t\) represents the improvement in fidelity, \(a\) and \(b\) are scaling coefficients, and \(p\) controls the type of scaling (e.g., \(p=1\) for linear, \(p=2\) for quadratic, etc.). By integrating the dynamics of the control hardware and atom qubit system into the environment's simulation, the RL agent learns to optimize the control sequence using the PPO algorithm from \textit{Stable Baselines3}. 
The code workflow is summarized in Algorithm~\ref{alg:ppo}.
More details are given in Appendix~\ref{appendix:ppo_rl}.

\begin{algorithm}
\caption{PPO-Based RL Approach}\label{alg:ppo}
\begin{algorithmic}[1]
\Require Initial policy parameters \(\theta\), value function parameters \(\phi\), old policy parameters \(\theta_{\text{old}} \leftarrow \theta\), environment \textit{QOCEnv}, reward scaling coefficients \(a\), \(b\), scaling exponent \(p\), PPO hyperparameters (clipping parameter \(\epsilon\), number of epochs \(K\), mini-batch size \(M\), total timesteps per update \(T\)), termination criteria.
\For{each iteration}
    \State \textbf{Collect Trajectories:} Run policy \(\pi_{\theta}\) in \textit{QOCEnv} for \(T\) timesteps to obtain trajectories \(\tau = \{(s_t, a_t, r_t, s_{t+1})\}\).
    \State \textbf{Compute Rewards:} For each timestep \(t\), compute the reward:
    \[
    r_t = a\, F_t^{p} + b\, \Delta F_t,
    \]
    where \(F_t\) is the fidelity at time \(t\) and \(\Delta F_t\) is the fidelity improvement.
    \State \textbf{Estimate Advantages:} Compute advantage estimates \(\hat{A}_t\) using the value function \(V_\phi(s_t)\) and the empirical returns.
    \State \textbf{Policy Update:}
    \For{\(k = 1, \ldots, K\)}
        \For{each mini-batch of size \(M\) sampled from \(\tau\)}
            \State Compute the probability ratio:
            \[
            r_t(\theta) = \frac{\pi_{\theta}(a_t\mid s_t)}{\pi_{\theta_{\text{old}}}(a_t\mid s_t)}
            \]
            \State Form the clipped surrogate objective:
            \[
            L^{\text{CLIP}}(\theta) = \mathbb{E}_t\Biggl[\min\Bigl(r_t(\theta)\hat{A}_t,\; \text{clip}\bigl(r_t(\theta),1-\epsilon,1+\epsilon\bigr)\hat{A}_t\Bigr)\Biggr]
            \]
            \State Update policy parameters \(\theta\) via gradient ascent on \(L^{\text{CLIP}}(\theta)\).
            \State Update value function parameters \(\phi\) by minimizing the squared error between \(V_\phi(s_t)\) and the empirical returns.
        \EndFor
    \EndFor
    \State \textbf{Update Old Policy:} Set \(\theta_{\text{old}} \leftarrow \theta\).
    \State \textbf{Termination Check:} If the target fidelity is reached or if performance stagnates, \textbf{break} the loop.
\EndFor
\State \Return Trained policy \(\pi_{\theta}\).
\end{algorithmic}
\end{algorithm}

\subsection{End-to-End Differentiable RL Approach}
\noindent Conventional RL methods often require extensive reward engineering and numerous training steps, making them inefficient for solving co-designed QOC problems. To address these challenges, we adopt an end-to-end gradient-based RL method~\cite{Singh2019EndtoEndRR} that directly minimizes the quantum gate error by backpropagating gradients through a differentiable simulation framework. In our approach, the control signal set \( \{V(t)\} \) is generated by a policy network implemented as a multilayer perceptron (MLP). The network maps a latent noise vector \( z \) (sampled from a standard normal distribution) through two fully connected layers (each with 64 neurons and \textit{tanh} activations) to produce an output matrix of dimensions \((N_{\text{ch}}, N_{\text{seg}})\), where \(N_{\text{ch}}\) is the number of independent control channels and \(N_{\text{seg}}\) is the number of piecewise segments used to approximate the continuous control signal. The optimization objective is to minimize the quantum gate infidelity, i.e., the cost function \(\mathcal{C}_f\) defined in Equ.~\ref{eq:optimization_problem}, with gradients \(\frac{\partial \mathcal{C}_f}{\partial V_i(t_k)}\) computed via automatic differentiation. Here, \(t_k\) denotes the discrete time points at which the control voltages are updated. The control parameters are updated using the gradient descent rule:
\begin{equation}
    V_i(t_k) \leftarrow V_i(t_k) - \eta \frac{\partial \mathcal{C}_f}{\partial V_i(t_k)},
\end{equation}
where \(\eta\) is the learning rate. This approach eliminates the need for stochastic policy-gradient updates as in PPO-based methods, and instead, optimizes the voltage sequences in a fully differentiable manner. To further improve convergence, curriculum learning is employed by starting with a coarse time grid and progressively increasing the resolution. The overall training process is summarized in Algorithm~\ref{alg:end2end}, with detailed implementation given in Appendix~\ref{appendix:end2end_rl}.

\begin{algorithm}
\caption{End-to-End Differentiable RL Approach}
\label{alg:end2end}
\begin{algorithmic}[1]
\State \textbf{Input:} Target unitary \(U_{\text{target}}\), initial policy parameters \(\theta\), number of phases \(P\)
\State \textbf{Initialize:} Random seed, policy network (MLP) with output dimension \((N_{\text{ch}}, S)\), where \(S\) is the number of piecewise segments.
\For{phase \(p=1,\dots,P\)}
    \State Set time resolution: update number of time steps \(T_p\) and segments \(S_p\)
    \If{\(p > 1\)}
        \State Update the policy network to increase resolution to \(S_p\)
    \EndIf
    \For{episode \(e=1,\dots,E_p\)}
        \State \textbf{Sample:} latent vector \(z \sim \mathcal{N}(0,I)\)
        \State \textbf{Policy Action:} Generate control schedule \(V(t)=\text{MLP}_{\theta}(z)\)
        \State \textbf{Environment Interaction:} Feed \(V(t)\) into the differentiable simulation \(\mathcal{S}\) and obtain \(U_{\text{sim}} = \mathcal{S}(V(t))\)
        \State \textbf{Reward Evaluation:} Compute reward \(r = F\big(U_{\text{sim}}, U_{\text{target}}\big)\)
        \State \textbf{Loss Calculation:} Define loss \(\mathcal{L} = 1 - r\)
        \State \textbf{Gradient Computation:} Compute \(\nabla_{\theta}\mathcal{L}\) via automatic differentiation
        \State \textbf{Parameter Update:} Update \(\theta \leftarrow \theta - \eta\, \nabla_{\theta}\mathcal{L}\) using an optimizer (e.g., Adam) with gradient clipping
        \State \textbf{Early Stopping Check:} If \(r\) exceeds a predefined threshold or improvement stagnates, break
    \EndFor
\EndFor
\State \textbf{Output:} Optimized policy parameters \(\theta^*\) yielding high-fidelity control schedules
\end{algorithmic}
\end{algorithm}

\section{Numerical Results and Discussions}
\label{sec:results}
\noindent We now present numerical results demonstrating the functionality of the hardware co-designed QOC framework. In Sec.~\ref{subsec:compqoc}, we first illustrate the necessity of employing QOC for mitigating hardware-related control imperfections by comparing traditional Gaussian-shaped pulses and optimal piecewise-shaped pulses obtained with the classical SADE-Adam optimizer, considering a simple program instruction with \(N_a\)=1. Then in Sec.~\ref{subsec:benchmark}, we benchmark the performance of three optimization methods, taking the classical hybrid SADE-Adam optimizer as a baseline and comparing conventional PPO-based RL optimizer and the End-to-End differentiable RL-based optimizer, assuming static control imperfections including inter-channel crosstalk and beam leakage based on models described in Sec.~\ref{sec:math_formulation}. We then further validate the performance robustness of the End-to-End differentiable RL-based optimizer considering randomized dynamic control imperfections. Training for RL-based methods was conducted on a single \textit{NVIDIA A30 GPU}, while the classical SADE-Adam optimizer was evaluated on a 16-core \textit{Intel Xeon CPU}. In all simulations, following parameters are used: laser intensity of \(20 \ mW/cm^2\), detuning of \(1 \ GHz\), hyperfine splitting of \(6.835 \ GHz\), fixed gate duration of \(0.1 \mu s\), and a total of 100 numerical time steps. 

\subsection{Necessity of QOC}
\label{subsec:compqoc}
\noindent In this section, we demonstrate that under non-ideal control conditions, a hardware co-designed QOC approach is essential for achieving high-fidelity gate operations. As \\ 

\newpage
\onecolumngrid
\begin{figure*}[!htbp]
    \includegraphics[width=0.99\textwidth]{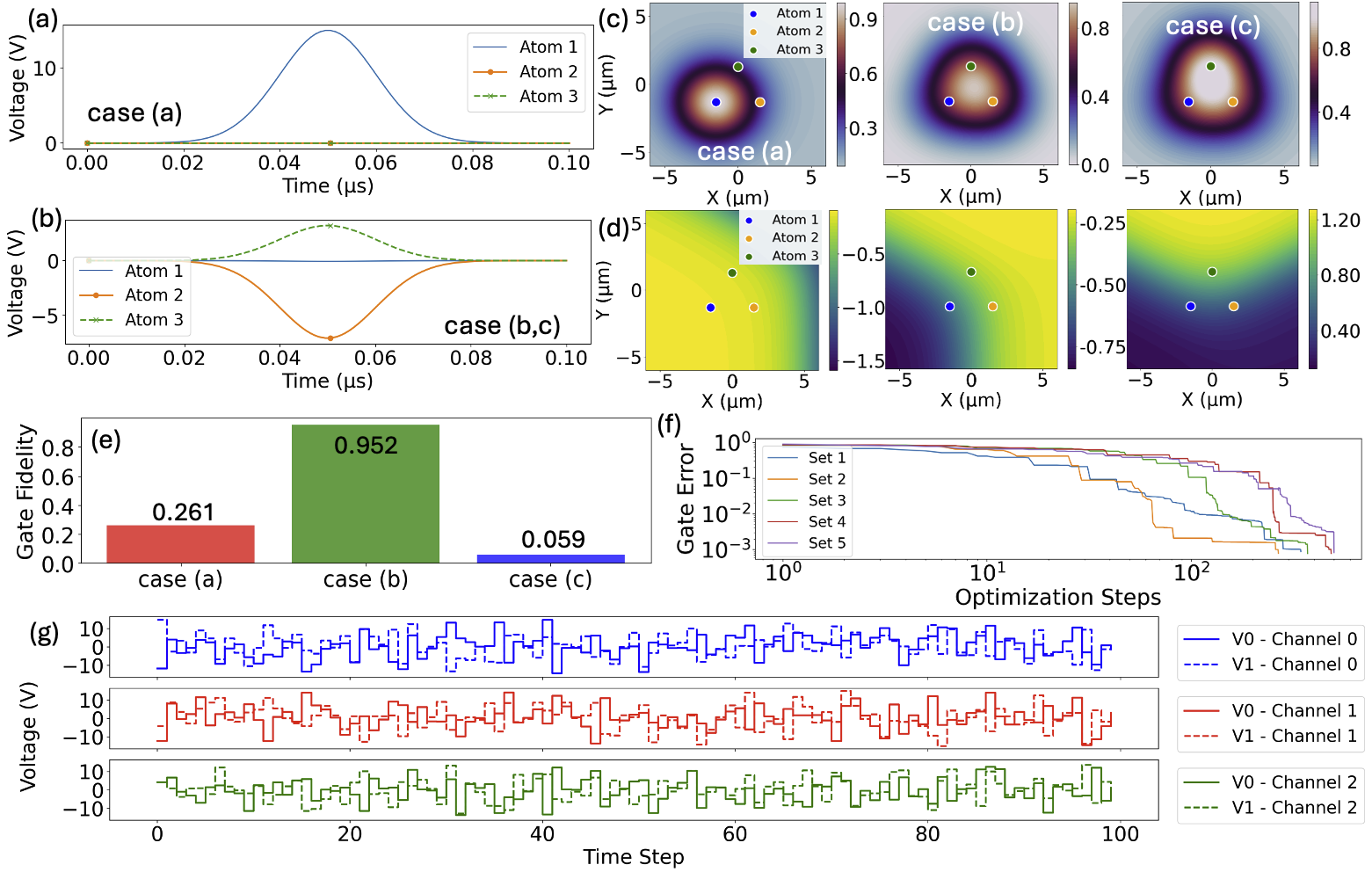}
    \caption{(a) Gaussian control pulse \(V(t)\) applied only to channel 1 targeting atom 1, aiming to implement the gate \(X_1/I_2/I_3\) on atoms 1, 2, and 3.
    (b) Optimized Gaussian control pulses \(V(t)\) applied to all channels, mitigating beam leakage and inter-channel crosstalk to implement the same target gate as in (a).
    (c, d) Normalized amplitude (c) and phase (d) profiles of the optical field on the 2D atomic plane, taken at half the total gate operation time. Field profiles for cases (a), (b), and (c) are shown from left to right. The inter-atomic spacing is set to 3~\(\mu\)m, and the beam waist is 2~\(\mu\)m, resulting in significant leakage between atoms.
    (e) Histogram comparing gate fidelities achieved in cases (a), (b), and (c), highlighting the fidelity degradation due to beam leakage and crosstalk.
    (f) Gate error optimization progress curves for the same target gate \(X_1/I_2/I_3\), using the classical hybrid SADE-Adam QOC optimizer. Each curve represents an independent test run initialized with a different random guess for the control pulses \(V_{\text{ini}}(t)\).
    (g) Example of optimized control pulses \(V(t)\) for channels 1–3 targeting atoms 1–3, from the first test run in (f). Control voltages are constrained within \([-15, +15]\) V, consistent with practical hardware limits.}
    \label{fig:compqoc}
\end{figure*} 
\twocolumngrid

\noindent an example, we consider a easy task, where the program instruction is to apply X-gate only on atom 1, and we can construct \( U_{\text{target}} = X_1 \otimes I_2 \otimes I_3 \). In an ideal scenario without inter-channel crosstalk or beam leakage, applying a Gaussian \(\pi\)-pulse exclusively to atom 1 while keeping the control signals for atoms 2 and 3 at zero, as shown in Fig.~\ref{fig:compqoc} (a), would implement \( U_{\text{target}} \) with high fidelity, as the control field would not influence neighboring atoms.
However, when atoms arranged in a triangular subgroup with a spacing of 3~\textmu m and a beam waist of 2~\textmu m, the control field applied to atom 1 inevitably leaks onto atoms 2 and 3, as illustrated in the left plots of the field's amplitude and phase profiles in Fig.~\ref{fig:compqoc} (c) and (d). In this case, noted as case (a), applying a Gaussian control pulse solely to atom 1 , results in a significantly low fidelity of \(\sim 0.262\), as seen in the first histogram bar in Fig.~\ref{fig:compqoc} (e).
To mitigate this undesired beam leakage effect, compensatory control pulses must be applied to atoms 2 and 3 to preserve their identity gates. In this case, noted as case (b), with optimized Gaussian control signals shown in Fig.~\ref{fig:compqoc} (b) applied to all atoms, 
\noindent the fidelity improves considerably to \(\sim 0.952\), as indicated by the second histogram bar in Fig.~\ref{fig:compqoc} (e). The corresponding field profiles in this case are given by the middle plots in Fig.~\ref{fig:compqoc} (c) and (d), where both amplitude and phase distributions change significantly compared with case (a).
However, when further including inter-channel crosstalk due to waveguide coupling, additional distortions of control fields profiles are induced, see the right plots in Fig.~\ref{fig:compqoc} (c) and (d), compared with middle plots for case (b). In this case, noted as case (c), the gate fidelity drops dramatically down to \(\sim 0.059\), as shown in third histogram bar in Fig.~\ref{fig:compqoc}(e). While additional fine-tuning of control pulses for atoms 2 and 3 could partially mitigate these additional field distortions, the complexity of the interference patterns induced by the inter-channel crosstalk grows as the control system scales up, where simple reshaping of Gaussian control pulses becomes very limited to counteract intricate crosstalk effects in a systematic way.
Instead, these challenges can be effectively addressed using the hardware co-designed QOC approach, which optimizes the set of control signals \( \{V(t)\} \) to implement the desired \( U_{\text{target}} \). To demonstrate this, we present results obtained using the  classical hybrid SADE-Adam optimizer, which iteratively refines the control signals to maximize fidelity. The inter-channel crosstalk and beam leakage are considered the same as in the case (c) using Gaussian pulses.
The progress curves of fidelity optimization in Fig.~\ref{fig:compqoc} (f) show rapid fidelity improvements with all final fidelities above 99.9\% (errors below $1\text{e-}3$ shown in the plot) across multiple trials with different initial guesses of control signals \(\{V_{\text{ini}}(t)\}\), highlighting the effectiveness and robustness of this approach. The final optimized control signals from one representative trial (Set 1) are shown in Fig.~\ref{fig:compqoc} (g). The piecewise-shaped control signals applied to DRMZMs in channels 0–2 manipulate the fields addressing atoms 1–3 and \(V_{0,1}\) in each channel represent signals applied to the two rings in a DRMZM of the PIC.

\subsection{Performance Benchmarking}  
\label{subsec:benchmark}  
\noindent In this section, we benchmark the performance of two RL-based QOC optimizers against the classical SADE-Adam optimizer. While our previous results focused on an easy task using SADE-Adam, we now consider two more challenging tasks. As mentioned at the end of Sec.~\ref{sec:phys_model},the intermediate task is considered as 
\(
U_{\mathrm{target}} = R_1 \otimes R_2 \otimes I_3,
\)  
and the difficult task is defined as  
\(
U_{\mathrm{target}} = R_1 \otimes R_2 \otimes R_3.
\)  
We denote these tasks as case \(N_g = 2\) (intermediate) and case \(N_g = 3\) (difficult).
For each task, tests were conducted using the three optimizers described in Sec.~\ref{sec:methods}. The performance of the classical SADE-Adam optimizer serves as the baseline for comparison with the PPO-based RL and the End-to-End differentiable RL methods. Our performance metrics include:
1) optimization ability: quantified by the final achieved optimal fidelity i.e. minimal gate error, and 
2) training efficiency: measured by the total number of optimization iterations or training episodes required for convergence.
All tests incorporate control imperfections---specifically, static inter-channel crosstalk and beam leakage effects---as described in Sec.~\ref{subsec:compqoc}. For each case, five test sets were generated, with target gates \(R_i\) randomly chosen from the extended Clifford plus T-gate group for each atom. Tables~\ref{tab:target_gates_Na2} and \ref{tab:target_gates_Na3} summarize these test sets for cases \(N_g = 2\) and \(N_g = 3\), respectively.

\begin{table}[h]
\centering
\begin{tabular}{c | c | c | c } 
 \hline
  & Hybrid SADE-Adam & PPO-Based RL  & End-to-End RL  \\ 
 \hline
 Set 1 & \{S, H, I\} & \{T, X, I\} & \{S, ZH, I\} \\
 Set 2 & \{HSX, ZH, I\} & \{XS, I, ZHS\} & \{SH, H, I\} \\
 Set 3 & \{XS, HSY, I\} & \{YH, I, SH\} & \{X, I, ZHS\} \\
 Set 4 & \{Y, I, HX\} & \{I, H, SH\} & \{HSY, I, S\} \\
 Set 5 & \{I, X, ZH\} & \{I, SY, XHS\} & \{I, ZHS, YHS\} \\
 \hline
\end{tabular}
\caption{Randomly selected target gates for \(N_g = 2\) across different optimizers.}
\label{tab:target_gates_Na2}
\end{table}
\begin{table}[h]
\centering
\begin{tabular}{c | c | c | c } 
 \hline
  & Hybrid SADE-Adam & PPO-Based RL  & End-to-End RL  \\ 
 \hline
 Set 1 & \{HSZ, ZS, H\} & \{T, T, HSX\} & \{SH, YHS, YH\} \\
 Set 2 & \{YH, XS, T\} & \{YH, YH, XHS\} & \{YHS, SH, SZ\} \\
 Set 3 & \{SY, SY, HX\} & \{SY, X, H\} & \{YH, HSZ, HX\} \\
 Set 4 & \{XH, XHS, XS\} & \{X, T, HSZ\} & \{ZS, ZH, Z\} \\
 Set 5 & \{ZHS, ZHS, YS\} & \{T, HSY, XH\} & \{HS, YS, XH\} \\
 \hline
\end{tabular}
\caption{Randomly selected target gates for \(N_g = 3\) across different optimizers.}
\label{tab:target_gates_Na3}
\end{table} 

\noindent For each optimizer, the optimization progress curves were averaged over five independent test runs, yielding a single performance curve that shows the mean gate error as a function of training episodes. The shaded bands represent standard deviations across runs. The results for all three optimizers are shown in Fig.~\ref{fig:benchmarking}(a), with the left and right panels corresponding to intermediate (\(N_g = 2\)) and difficult (\(N_g = 3\)) task settings, respectively. The hybrid SADE-Adam optimizer is shown in red, the PPO-based RL in green, and the end-to-end differentiable RL method in blue.
For the intermediate task (\(N_g = 2\)), the SADE-Adam optimizer achieves the lowest final gate error (\(1.529\text{e}{-3} \pm 4.069\text{e}{-4}\)), outperforming both the end-to-end RL method (\(7.526\text{e}{-3} \pm 6.171\text{e}{-3}\)) and the PPO-based RL method (\(7.366\text{e}{-2} \pm 2.496\text{e}{-2}\)). Notably, the end-to-end RL still improves upon PPO-RL by nearly an order of magnitude in both final gate error and training efficiency.
As the task becomes more difficult (\(N_g = 3\)), the performance trends reverse. The end-to-end RL approach now achieves the best result, with a final gate error of \(8.234\text{e}{-4} \pm 2.579\text{e}{-4}\), outperforming SADE-Adam (\(1.090\text{e}{-3} \pm 3.946\text{e}{-4}\)) and significantly surpassing PPO-RL (\(9.820\text{e}{-2} \pm 3.192\text{e}{-2}\)). In this scenario, end-to-end RL reduces the gate error by more than two orders of magnitude compared to PPO-RL and converges substantially faster.
These results highlight a key distinction: while classical evolutionary methods like SADE-Adam perform well in moderately complex settings, their advantage diminishes as the system scales. The PPO-based RL method, although adaptive, struggles with scalability and suffers from instability in high-dimensional search spaces. In contrast, the differentiable end-to-end RL approach demonstrates strong scalability and adaptability, making it particularly well-suited for optimizing parallel quantum gate operations under realistic hardware constraints.

\subsection{Robustness Validation}
\noindent To further validate the robustness of the end-to-end differentiable RL-based QOC optimizer under realistic hardware conditions, we evaluate its performance in two additional challenging scenarios for the \(N_g = 3\) task:

\begin{enumerate}
    \item \textit{Case 1 – Static Inter-Channel Crosstalk Variation:} 
\newpage
\onecolumngrid
\begin{figure*}[!htbp] 
    \includegraphics[width=0.99\textwidth]{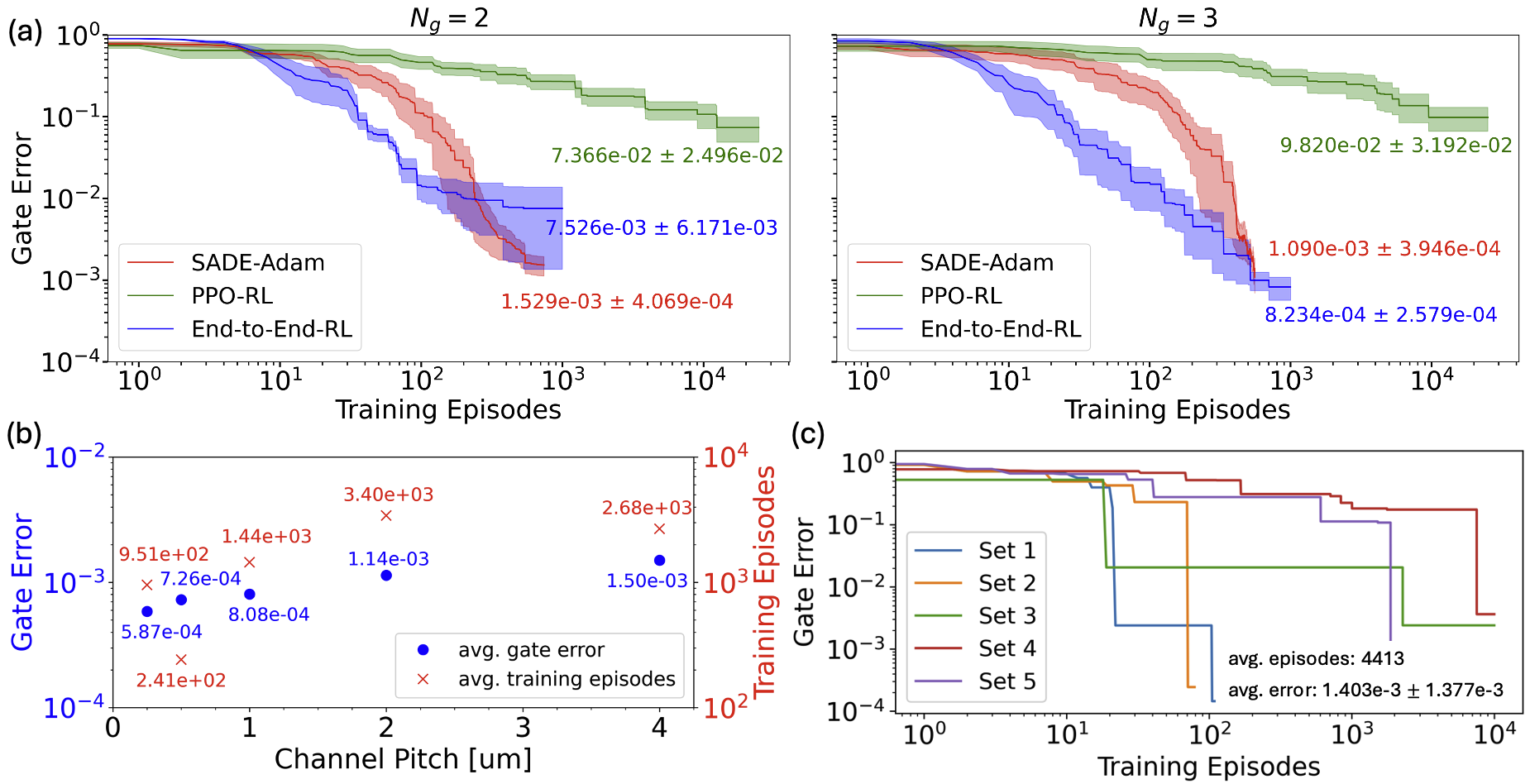}
    \caption{(a) Gate error optimization progress for intermediate and difficult tasks with \(N_g = 2\) (left) and \(N_g = 3\) (right), comparing the performance of the classical SADE-Adam optimizer, conventional PPO-based RL, and the proposed end-to-end differentiable RL-based QOC approach. Each curve shows the average over five test runs; solid lines indicate the mean, and shaded regions represent standard deviations. Static inter-channel crosstalk and beam leakage are assumed.
    (b) Optimized gate errors (blue dots, left y-axis) and corresponding training episodes (red crosses, right y-axis) achieved using the end-to-end RL-based QOC optimizer for different channel pitch values in the PIC, representing varying levels of static inter-channel crosstalk. Each point is averaged over five test runs with randomly selected target gate sets.
    (c) Gate error optimization progress for a difficult task with \(N_g = 3\), using the end-to-end RL-based QOC optimizer under dynamically varying control imperfections, where both inter-channel crosstalk and beam leakage fluctuate over time. Each curve corresponds to an independent test run with a randomly selected target gate set. }
\label{fig:benchmarking}
\end{figure*}
\twocolumngrid 
\noindent The beam leakage effect is fixed, while the strength of static inter-channel crosstalk in the PIC is varied to cover a wide range of fabrication-induced variations across different chips.    
    \item \textit{Case 2 – Dynamic Control Imperfections:} Both beam leakage and inter-channel crosstalk are made time-dependent to mimic temporal drifts and fluctuations in hardware performance during operation.
\end{enumerate} 
In Case 1, we vary the strength of static inter-channel crosstalk by sweeping the base channel pitch \(d_0\) as described in Sec.~\ref{sec:phys_model}. In Case 2, dynamic imperfections are introduced by applying time-varying perturbations to the nominal coupling and decay parameters:
\[
\kappa_{m,n}(t) = \kappa_{m,n}^0 + \delta \kappa_{m,n}(t), \quad \alpha_{m,n}(t) = \alpha_{m,n}^0 + \delta \alpha_{m,n}(t),
\]
where the perturbations are sampled at each time step \(t_i\) from uniform distributions:
\[
\delta \kappa_{m,n}(t_i) \sim \mathcal{U}(-\Delta\kappa,\, \Delta\kappa), \quad \delta \alpha_{m,n}(t_i) \sim \mathcal{U}(-\Delta\alpha,\, \Delta\alpha),
\]
with \(\Delta\kappa = 0.5\) and \(\Delta\alpha = 0.2\). Beam leakage fluctuations are modeled by allowing the beam waist to vary as:
\[
w_0(t) = w_0^0 + \delta_w(t), \quad \delta_w(t) \sim \mathcal{U}(-\Delta w,\, \Delta w),
\]
with \(\Delta w = 0.1\,\mu\text{m}\). These variations serve as stringent tests of the optimizer’s robustness under both static and dynamic hardware-induced distortions.

\medskip

\noindent The results are presented in Fig.~\ref{fig:benchmarking} (b) and (c). 
In Case 1, Fig.~\ref{fig:benchmarking} (b) shows the optimized final gate error (blue dots, left y-axis) and the corresponding number of training episodes (red crosses, right y-axis), each averaged over five independent test runs for different values of the channel pitch (\(0.25\,\mu\text{m}\), \(0.5\,\mu\text{m}\), \(1.0\,\mu\text{m}\), \(2.0\,\mu\text{m}\), and \(4.0\,\mu\text{m}\)). Each run uses a randomly selected target gate set (see Table~\ref{tab:target_gates_Na3_varyPitch}, Case 1).
Even in the worst-case crosstalk scenario, the optimized gate error remains on the order of \(1\text{e}{-3}\), demonstrating the robustness of the end-to-end differentiable RL-based optimizer. Interestingly, as the channel pitch decreases, which corresponds to stronger inter-channel crosstalk, both the gate error and the number of training episodes tend to decrease. This counterintuitive trend suggests that moderate crosstalk may actually facilitate multi-channel coordination, thereby enhancing the ability to implement parallel single-qubit gates using local addressing beams.
In Case 2, Fig.~\ref{fig:benchmarking} (c) presents the gate error optimization progress for five independent test runs, each using a randomly selected target gate set (see Table~\ref{tab:target_gates_Na3_varyPitch}, Case 2). The average final gate error is \(1.403\text{e}{-3} \pm 1.377\text{e}{-3}\), which remains comparable to the worst-case result observed in Case 1, despite the additional complexity introduced by dynamic imperfections.
However, the average number of training episodes required increases by approximately 1000 compared to the best-case static crosstalk scenario in Case 1. This increase reflects the added difficulty of optimizing control under randomized, time-varying distortions in both inter-channel crosstalk and beam leakage. These results further confirm that the end-to-end differentiable RL-based optimizer maintains strong robustness even under temporally fluctuating hardware conditions.

\begin{table}[!htbp]
\centering
\begin{tabular}{c | c | c } 
 \hline
  & Case 1 & Case 2 \\ 
 \hline
 Set 1 & \{HSY, S, ZS\} & \{HSX, YHS, ZS\}  \\
 Set 2 & \{SH, YHS, YH\} & \{SH, SH, ZS\} \\
 Set 3 & \{ZS, SY, T\} & \{ZS, SH, ZH\}  \\
 Set 4 & \{Z, YH, S\} & \{YHS, HX, YH\}  \\
 Set 5 & \{SZ, SH, H\} & \{HX, SZ, Z\} \\
 \hline
\end{tabular}
\caption{Randomly selected target gate sets for \(N_g = 3\) on atoms \(\{1,2,3\}\), used to evaluate robustness under static crosstalk variations (Case 1) and dynamic control imperfections (Case 2).}
\label{tab:target_gates_Na3_varyPitch}
\end{table}

\section{Conclusion and Outlook}
\label{sec:conclusion}
\noindent In this work, we introduced a hardware co-designed quantum control framework that integrates photonic control hardware modeling with the QOC theory to address robust control challenges in atomic quantum processors. Specifically, we considered physical imperfections arising from inter-channel crosstalk induced by waveguide coupling in the programmable PIC and beam leakage due to imperfect optical steering from the SLM.
Using a classical hybrid SADE-Adam optimizer as a baseline, we demonstrated that a novel end-to-end differentiable RL-based QOC optimizer consistently outperforms both the classical and conventional PPO-based RL approaches. While PPO performance degrades as system complexity increases, the end-to-end method reliably achieves gate fidelities above 99.9\% and maintains strong robustness under both static and dynamic control imperfections.
These results underscore the practical potential of combining RL-based QOC techniques with detailed physical models of imperfect hardware. The proposed framework enables automated, high-fidelity control of parallel quantum gate operations in realistic settings, offering a scalable and generalizable solution for quantum engineering at scale.
Looking ahead, future work will extend this framework to include multi-qubit entangling gates and incorporate decoherence effects to better align with experimental conditions. Moreover, integrating this co-designed control framework with quantum compilers opens the door to hardware-aware quantum circuit synthesis and quantum error correction, ultimately enabling more scalable and fault-tolerant quantum computing architectures.

\section*{Data and Code Availability}
\noindent The source code, simulation results, and example notebooks for this work are publicly available on GitHub: 
\href{https://github.com/dingq1/rlqoc_codesign_apic}{\texttt{https://github.com/dingq1/rlqoc\_codesign\_apic}}.

\appendix
\section{Hybrid SADE-Adam Approach}
\label{appendix:sade_adam}
\noindent In this section, we explain details of the classical hybrid SADE-Adam optimizer, particularly the implementation of the gradient-free SADE and gradient-based Adam methods are given as below.
\\
\paragraph{Gradient-free global search}

SADE is employed in the first optimization stage to explore the high-dimensional, non-convex search space of control signals. It maintains a population of candidate solutions, denoted as \textit{popsize}, which evolves iteratively through mutation, crossover, and selection. Unlike standard differential evolution, SADE dynamically adjusts the mutation and crossover rates based on their past success, enabling an adaptive balance between exploration and exploitation.
In each generation \( g \), a new candidate solution \( \{V_i^{(g+1)}(t)\} \) for the \( i \)-th individual in the population is generated through mutation, which combines three randomly selected parent solutions \( \{V_a^{(g)}(t)\} \), \( \{V_b^{(g)}(t)\} \), and \( \{V_c^{(g)}(t)\} \) from the current population:
\begin{equation}
    \{V_i^{(g+1)}(t)\} = \{V_a^{(g)}(t)\} + \mu \cdot (\{V_b^{(g)}(t)\} - \{V_c^{(g)}(t)\}).
\end{equation}
Here, \( g \) represents the generation index, \( i \) indexes individuals within the population, \( V_{a,b,c}^{(g)}(t) \) are three distinct candidates randomly chosen from the population, ensuring \( a, b, c \neq i \), and \( \mu \) is a mutation factor dynamically sampled between 0.1 and 0.9.
Following mutation, crossover is applied, where each component of \( \{V_i^{(g+1)}(t)\} \) is replaced with that of the original parent \( \{V_i^{(g)}(t)\} \) with a probability \( CR \), known as the crossover rate. This probability is adaptively varied within the range \([0.1, 0.9]\), allowing the algorithm to adjust its search strategy based on prior success.
Selection ensures that the new candidate \( V_i^{(g+1)}(t) \) replaces \( V_i^{(g)}(t) \) only if it improves the optimization objective, which is to minimize the gate error. If the new solution does not lead to an improvement, the original candidate is retained. This process preserves a constant population size throughout the optimization.
SADE continues evolving solutions until either the maximum number of generations is reached or the fidelity surpasses a predefined threshold, set to 0.95 in this case. The best-performing candidate from SADE serves as the initial guess for the subsequent fine-tuning using gradient-based Adam optimization.
\\
\paragraph{Gradient-based local search} 
Adam optimization is used for fine-tuning the control signals after SADE. The parameters are initialized from the best solution found by SADE, and Adam iteratively refines them using gradients computed via automatic differentiation in JAX. The update rule follows:
\begin{equation}
    \{V_i(t)\} \leftarrow \{V_{i-1}(t)\} - \eta \frac{\{m_i(t)\}}{\sqrt{\{v_i(t)\} + \epsilon}},
\end{equation}
where \( i \) is the iteration index, and \( \{m_i(t)\} \) and \( \{v_i(t)\} \) are the first- and second-moment estimates of the gradients for \( \{V_i(t)\} \). The term \( \epsilon \) is a small constant to prevent numerical instability when \( v_i(t) \) is close to zero.
The learning rate \( \eta \) is initialized at 0.0001 and is adaptively reduced when the fidelity exceeds predefined thresholds of 0.98, 0.99, 0.995, and 0.997, following decay factors of 0.5, 0.2, 0.5, and 0.2, respectively. This decay mechanism ensures stable convergence while preventing large parameter updates that may degrade fidelity.
Adam optimization continues until one of the following stopping criteria is met: the fidelity surpasses a predefined threshold 0.999, or the optimization shows no significant improvement for certain numbers of consecutive steps.

\section{PPO-based RL Approach}
\label{appendix:ppo_rl}
\noindent In this section, we explain details of the implemented conventional PPO-based RL approach as QOC optimizer, including the working principle of PPO, construction of a custom RL environment, employed neural network architecture and training strategies. \\

\paragraph{Working Principles of PPO} 
PPO optimizes a stochastic policy \(\pi_\theta(a \mid s)\) parameterized by \(\theta\), where \(s\) represents the environment state and \(a\) represents the control action. The objective in our QOC problem is to maximize the expected cumulative reward associated with improving quantum gate fidelity over multiple time steps:
\begin{equation}
J(\theta) = \mathbb{E}_{\pi_\theta} \left[ \sum_{t=0}^{T} \gamma^t r_t \right],
\end{equation}
where \(r_t\) represents the reward at time step \(t\), and \(\gamma\) is the discount factor. The reward function is designed to encourage both immediate fidelity improvements and long-term optimal control performance. PPO updates the policy using a clipped surrogate objective function to ensure stable learning:
\begin{equation}
L^\text{PPO}(\theta) = \mathbb{E} \left[ \min \left( r_t(\theta) A_t, \text{clip}(r_t(\theta), 1-\epsilon, 1+\epsilon) A_t \right) \right],
\end{equation}
where \(r_t(\theta)\) is the probability ratio between the new and old policies, \(A_t\) is the advantage function, and \(\epsilon\) is the clipping parameter (typically set to \(0.2\)). In our QOC framework, this prevents excessive changes in the control signals, ensuring a smooth learning process and avoiding sudden variations that may degrade gate fidelity.
\\
\paragraph{Custom RL Environment for QOC}
To apply PPO to the QOC problem, we construct a custom Gymnasium environment called \textit{QOCEnv}, where the agent optimizes the control signal set \(\{V(t)\}\) that programs the light fields applied to atoms to implement the gate operation and thereby determine the gate fidelity. The state consists of the history of control voltages over past time steps along with the current fidelity progress. The action space comprises continuous voltage adjustments in the range \([-1, 1]\), which are later scaled to the physical range of \([-15\,\mathrm{V}, 15\,\mathrm{V}]\). The reward function is carefully tailored to incentivize fidelity improvement via adaptive scaling:
\begin{equation}
r_t = a\, F_t^{p} + b\, \Delta F_t,
\end{equation}
where \(F_t\) denotes the fidelity at time \(t\), \(\Delta F_t\) represents the improvement in fidelity, \(a\) and \(b\) are scaling coefficients, and \(p\) controls the type of scaling (e.g., \(p=1\) for linear, \(p=2\) for quadratic, etc.). To further improve training stability, the reward scaling mechanism adapts based on the fidelity progress, ensuring effective learning across different fidelity regimes and enabling early termination if fidelity stagnates.
\\
\paragraph{Feature Extraction and Policy Architecture}
The PPO agent follows an actor–critic architecture, where the actor selects control signals based on the current state, and the critic evaluates the expected return of that state. In our QOC problem, the control signal set \(\{V(t)\}\) comprises time sequences that are highly correlated across adjacent time steps. To exploit this structure, the network architecture includes convolutional layers~\cite{lecun2015deep}, which capture how previous voltage values influence current fidelity, smooth transitions in control signals to reduce noise, and enhance sample efficiency by reducing the number of parameters. The network consists of an input layer that accepts the history of voltage sequences and the current fidelity progress, followed by a convolutional feature extractor. Our implementation employs two 2D convolutional layers (with 32 and 64 filters, respectively, a kernel size of 3, and ReLU activation) and a fully connected layer with 256 neurons. Finally, the network splits into two branches: the actor network outputs the mean and variance of the Gaussian policy \(\pi_\theta(a \mid s)\) to determine voltage adjustments, while the critic network estimates the value function \(V(s)\) to predict expected fidelity improvement. Training is parallelized across multiple environments using \textit{SubprocVecEnv} in \textit{Stable Baselines3}, enabling efficient exploration of the control landscape.
\\
\paragraph{Training Strategy and Termination Criteria}
The PPO agent is trained over 10,000 episodes with a learning rate of \(1 \times 10^{-4}\), an entropy coefficient of \(0.1\) to encourage exploration, a discount factor of \(\gamma = 0.99\), a clipping threshold of \(0.2\), generalized advantage estimation (GAE) with \(\lambda = 0.95\), and a maximum gradient norm of \(0.5\) for gradient clipping. During training, the RL-based optimizer iteratively refines the control signals to maximize gate fidelity. An adaptive logging system tracks fidelity improvements and saves the corresponding voltage sequences whenever a new maximum fidelity is achieved. Training terminates when either the fidelity reaches \(F \geq 0.999\) or if the best fidelity stagnates for over 1000 episodes within the range \(0.99 \leq F < 0.999\). Additionally, the adaptive reward scaling mechanism can detect prolonged stagnation and trigger early termination.

\section{End-to-End Differentiable RL Approach}
\label{appendix:end2end_rl}
\noindent In this section, we explain the implementation of the end-to-end differentiable RL-based optimizer, including the basic differentiable treatment of the co-designed QOC problem, employed policy network architecture and curriculum learning based training strategies. 
\\
\paragraph{End-to-End Backpropagation Through Time}
In this formulation, the control policy is defined as a continuous function \( \{V(t)\} = \{V_1(t), V_2(t), \dots, V_{N_{\text{ch}}}(t)\} \), where each \( V_i(t) \) represents the time-dependent voltage signals applied to the two rings of the control unit DRMZM in a specific PIC channel. The control signals are parameterized by a neural network that outputs an \( N_{\text{ch}} \times N_{\text{seg}} \) matrix, where \( N_{\text{ch}} \) denotes the number of independent control channels, and \( N_{\text{seg}} \) represents the number of discrete time segments used to approximate the continuous control sequence.
The optimization objective is to minimize the quantum gate infidelity, see cost function \( \mathcal{C}_f \) given in Equ.~\ref{eq:optimization_problem} , with gradients \( \frac{\partial \mathcal{C}_f}{\partial V_i(t_k)} \) computed directly via automatic differentiation . Here, \( t_k \) represents the discrete time points where the control voltages are updated, with \( k \) indexing the segments in the piecewise discretization of the control sequence. The control parameters are updated using the gradient descent rule:
\begin{equation}
    V_i(t_k) \leftarrow V_i(t_k) - \eta \frac{\partial \mathcal{C}_f}{\partial V_i(t_k)},
\end{equation}
where \( \eta \) is the learning rate. This eliminates the need for stochastic policy-gradient updates as in PPO-based RL approach and instead optimizes the voltage sequences in a fully differentiable manner.
\\
\paragraph{Neural Network Policy Architecture}
The policy network is a minimal yet expressive multilayer perceptron (MLP) designed to generate correlated and structured control sequences across all channels. It consists of two fully connected layers with 64 neurons each, using \textit{tanh} activation functions to enforce smooth control variations. The input to the network is a latent noise vector \( z \), sampled from a standard normal distribution, which enables the model to explore a diverse set of control strategies. The network transforms this latent input into structured control voltages using a hierarchical mapping.
The first layer applies a non-linear transformation to the latent vector, capturing global correlations in the control sequence. The second layer refines these representations, ensuring that the output follows physically meaningful constraints. The final output layer produces an \( N_{\text{ch}} \times N_{\text{seg}} \) matrix, where each element represents the control voltage for a specific channel at a given time segment. The output values are linearly scaled to match the predefined voltage range \( [-15V, 15V] \), fulfilling the hardware imposed constraints.
This formulation allows the policy network to learn temporally and spatially correlated control sequences, optimizing the overall quantum evolution trajectory.
\\
\paragraph{Curriculum Learning-Based Training Strategy}
To improve convergence stability and efficiency, we employ a curriculum learning strategy~\cite{curriculumLR} where control sequences are optimized progressively with increasing time resolution. Training begins with a coarse discretization at \( t_{\text{steps}} = 20 \), allowing the optimizer to explore the control landscape efficiently and identify approximate solutions that significantly improve gate fidelity. Once fidelity stabilizes, the resolution increases to \( t_{\text{steps}} = 50 \), refining control sequences with better temporal resolution and smoother transitions.
The final stage optimizes at \( t_{\text{steps}} = 100 \), achieving precise control over gate implementation. At each stage, the optimizer updates the control voltage sequences using gradient descent. The optimized control from the previous stage initializes the next phase, ensuring smooth progression rather than restarting from scratch. This gradual refinement prevents instability and enables hierarchical learning of optimal control sequences. Training continues until fidelity surpasses \( F \geq 0.999 \) or improvement stagnates for multiple iterations, triggering adaptive learning rate decay. If the maximum number of iterations is reached, training terminates automatically.

\section*{Acknowledgments}
\noindent Qian Ding acknowledge funding support from the \textit{Swiss National Science Foundation (SNSF) Postdoc Mobility Fellowship}. 
Dirk Englund acknowledge funding from \textit{NSF CQN}. 
We acknowledge the support from the \textit{NVIDIA Academic Grant} and \textit{subMIT} cluster at MIT for offering access to high-performance computational resources.

\section*{References}
\bibliographystyle{unsrt}
\bibliography{references}  

\end{document}